\documentclass[prl,twocolumn,shownopacs]{revtex4}
\usepackage{graphicx,epsfig}
\usepackage{times,amsmath,amssymb,latexsym}
\usepackage{rotating}
\usepackage{stmaryrd}

\newcommand{\eqn}[1]{\begin{eqnarray} #1 \end{eqnarray}}

\newcommand{\ci}{{\rm i}}

\newcommand{\twoDmatrix}[4]{
             \left(\begin{array}{cc} #1&#2\\ #3&#4 \end{array}\right)}

\begin{document}

\title{Cold atoms in non-Abelian gauge potentials:\\ From
the Hofstadter ``moth'' to  lattice gauge theory}

\author{K. Osterloh$^1$, M. Baig$^2$, L. Santos$^3$,
P. Zoller$^4$, and M. Lewenstein$^{1,5}$}

\affiliation{\mbox{(1)~Institut~f\"ur~Theoretische~Physik,~Universit\"at~Hannover,~D-30167~Hannover,~Germany}\\
\mbox{(2)~IFAE, ~Universitat~Aut\`onoma~de~Barcelona,~08193~Bellaterra~(Barcelona),~Spain}\\
\mbox{(3)~Institut~f\"ur~Theoretische~Physik~III,~Universit\"at~Stuttgart,~D-70550~Stuttgart}\\
\mbox{(4)~Institute~for~Theoretical~Physics,~University~of~Innsbruck,~A-6020~Innsbruck,~Austria}
\mbox{(5)~ICFO~---~Institut~de~Ci\`{e}ncies~Fot\`{o}niques,~08034~Barcelona,~Spain}}
\begin{abstract}
We demonstrate how  to create artificial external non-Abelian gauge potentials
acting on cold atoms in optical lattices.
The method employs $n$ internal states of atoms and laser assisted state
sensitive tunneling. Thus, dynamics are communicated by unitary
$n\times n$-matrices. By experimental control of
the tunneling parameters, the system can be made truly non-Abelian.
We show that single particle dynamics in the case of intense $U(2)$
vector potentials lead to a generalized Hofstadter butterfly spectrum
which shows a complex ``moth''-like structure. We discuss the possibility
to employ non-Abelian interferometry (Aharonov-Bohm effect)
and address methods to realize matter dynamics in specific classes of
lattice gauge fields.
\end{abstract}

\date{\today }
\maketitle




One of the most significant trends in the physics of ultra-cold gases
nowadays concerns, without any doubts,  strongly correlated systems. Apart
from systems at BCS-BEC
crossover, low dimensional
systems, and atomic lattice gases, perhaps the most fascinating
possibilities are offered by atomic gases subject to the effects of artificial
magnetic fields. A feasible realization,
albeit experimentally challenging, is an atomic system in a rotating trap.
Trap rotation mimics the effects of external homogeneous
magnetic fields. In the situation, when the rotation frequency
approaches the trap frequency, the systems may allow to study the fractional
quantum Hall effect (FQHE) and Laughlin liquids \cite{CiracZoller}.
Several experiments  with gases in rotating traps are currently  underway
\cite{Dalibard}. In this context, rotating dipolar lattice gases
(for a review, see \cite{nobel}) offer new possibilities and
advantages compared to gases with short ranged interactions, as
the energy gap between the Laughlin
state and quasi-hole excited states survives the large $N$ limit. This
allows for the realization of Laughlin
liquids with a few hundred atoms at realistic temperatures \cite{klaus}.
Very recently, the creation of artificial magnetic fields in terms of
electro-magnetically induced transparency (EIT) has been proposed
in Ref.~\cite{patrick}. 

\begin{figure}[tbp]
\centering
\caption{The Hofstadter butterfly.
Eigenenergies in natural units plotted versus magnetic flux per plaquette
$\alpha=p/q,\in[0,\, 1]$, with $q\leq 100$. The lowest Bloch band splits into
subbands and forms a fractal for the set of all $\alpha$'s. }
\includegraphics[width=0.45\textwidth]{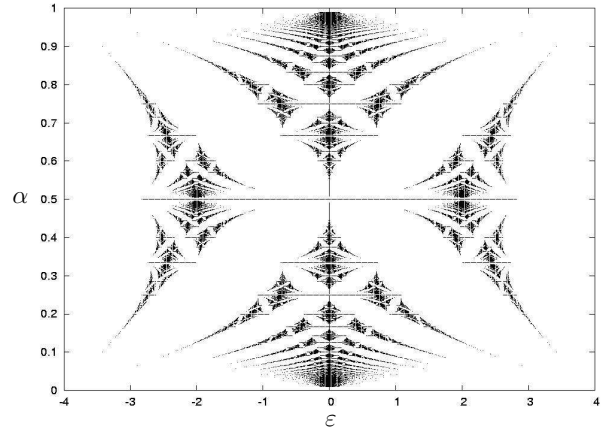}
\label{Hof}
\vspace*{-0.5cm}
\end{figure}

Alternatively, magnetic field effects can be realized in a lattice
by introducing appropriate phase factors for tunneling amplitudes.
If multiplied around an elementary plaquette, these lead to phases proportional
to the magnetic flux penetrating the plaquette.
Jaksch and Zoller, and subsequently  other groups,
have recently proposed methods to realize such ``artificial'' magnetic
field effects in lattice gases. These employ atoms with multiple internal
states, laser assisted tunneling, lattice tilting (acceleration), and other
experimentally accessible techniques \cite{jakschbut,mueller,demler}.
The physical features in such ``artificial'' magnetic fields are
extremely rich. For single atoms, the
 spectrum exhibits the fractal Hofstadter
``butterfly'' structure\cite{hofstadter} illustrated in Fig.~\ref{Hof}.
In weakly interacting or
weakly disordered  systems, the modifications of the
butterfly due to interactions and disorder, respectively,
may be studied \cite{jakschbut}.
Finally, considering strong magnetic fields in the limit of strong interactions,
Laughlin like states are expected \cite{demler}.
In this Letter, we show that the use  of atoms with more internal states,
the application  of state dependent, laser assisted tunelling
(cf. \cite{wilczek}), and coherent transfer between
internal states, allows for a generalization of the above methods and
creation of ``artificial external magnetic fields''
 corresponding to non-Abelian $U(n)$,
 or $SU(n)$ gauge fields \cite{equiv}.
In this case, tunneling amplitudes are replaced by
unitary matrices whose product
around a plaquette is non-trivial and its mean trace (Wilson loop)
is not equal to $n$ \cite{gauge}.
To our  knowledge, the physics of matter
in designed non-Abelian gauge fields has not been studied at all in the AMO
physics context. Although single particles in
$SU(n)$ monopole fields have been intensively studied in high energy physics
\cite{monopole}, other field configurations have not attracted such interest.
Specific examples of external gauge fields have also been discussed with
respect to NMR, nuclear and  molecular physics \cite{shapere}. 
Recently, there has been a proposal to create non-Abelian gauge fields 
using EIT
\cite{fleisch}. However, all of
these proposals either do not deal with interacting many particle systems in such
gauge fields, or they do not consider lattice gauge fields of ultra-high
strength.

Having introduced the designed external non-Abelian gauge fields in
optical lattices,  we study their influence on single particle dynamics.
We generalize the Hofstadter butterfly to the non-Abelian case
and show that in case of intense $U(2)$ gauge fields,
the spectrum (as a function of two ``magnetic fluxes'') exhibits a complex
``moth'' structure of holes. We discuss the possibility to observe
the $U(2)$ Aharonov-Bohm effect, to realize non-Abelian interferometry,
and to simulate lattice gauge dynamics using random field configurations.

We start from the physics of a single particle in a two-dimensional
square lattice of spacing $a$ in the presence of an Abelian magnetic
field $B$. When the
lattice potential is  sufficiently strong, the single particle
non-relativistic Hamiltonian in the tight binding approximation is given by the
so-called Peierl's substitution \cite{hofstadter},
and reads
\begin{equation}
H\!=\!-2V_0\!\!\left[
     {\rm cos}\Big(\frac{a}{\hbar}(p_x\!-\ci\frac{{\rm e}}{c}A_x)\!\Big)
     \!+\!{\rm cos}\Big(\frac{a}{\hbar}(p_y\!-\ci\frac{{\rm e}}{c}A_y)\!\Big)
     \!\right]\!,\label{H}
\end{equation}
where $V_0$ is the optical potential strength,
$\vec{p}$ is the momentum operator, and $\vec{A}$ is the magnetic vector
potential.  Given $[p_j,A_j]=0$,
the single-particle wave equation reads
\eqn{\label{spwaveequation}
      e^{-\ci\frac{{\rm e}a}{\hbar c}A_x}\psi(x\!+\!a,y)
     +e^{\ci\frac{{\rm e}a}{\hbar c}A_x}\psi(x\!-\!a,y)&&\\
     +e^{-\ci\frac{{\rm e}a}{\hbar c}A_y}\psi(x,y\!+\!a)
     +e^{\ci\frac{{\rm e}a}{\hbar c}A_y}\psi(x,y\!-\!a)
     &\!\!=\!\!&-\frac{E}{V_0}\psi(x,\,y)\nonumber}
The choice of $\vec{A}$  determines the magnetic field $\vec B$, which in turn
determines the behavior of the system.
For $\vec B=B\vec{\rm e}_z$,
one may choose  the potential $\vec{A}=(0,\,Bx,\,0)$, thus, 
solely tunnelings in the $y$-direction acquire phases.
This makes the problem effectively one-dimensional
and \eqref{spwaveequation} can be transformed into Harper's
equation \cite{harper}:
\eqn{g(m\!+\!1)-g(m\!-\!1)+2\cos(2\pi m\alpha-\nu)g(m)
=\varepsilon g(m)\label{HarperEquation}}
by using the ansatz $\psi(ma,\,na)={\rm e}^{\ci\nu n}g(m)$, where
$x=ma$, $y=na$, and $\varepsilon=-E/V_0$.
Solving the eigenvalue problem for specific,
i.e., rational values of the magnetic flux $\alpha=\frac{{\rm e}a^2B}{hc}$
per elementary plaquette makes the problem periodic.
This results in a band spectrum whose gaps form the famous
Hofstadter butterfly (Fig.~\ref{Hof}). Note that the regime
of this spectrum requires finite values of $\alpha$, i.e.,
magnetic fields $B\sim 1/a^2$, which in the continuum limit $a\to 0$
become ultra-intense.
A physical system satisfying \eqref{HarperEquation} can be realized
in a simpler way in a 1D optical lattice setup with standard tunneling
and an overlaid super-lattice potential with spatial period $1/\alpha$.
This statement also holds for the non-Abelian generalizations
in the scope of this manuscript. 
Anyhow, the full 2D realization proposed in \cite{jakschbut} allows for 
a more powerful control opening up completely novel features in such a system.
%
%

We consider an atomic gas in a  3D optical lattice and
assume that tunneling is completely suppressed in the $z$-direction,
so that, effectively, we deal with an array of 2D lattice gases
and are able to restrict ourselves to one copy.
The atoms occupy two
Internal hyperfine states $|g\rangle$, $|e\rangle$, and the
optical potential traps them  in the state $|g\rangle$ and $|e\rangle$,
respectively, in every second column, i.e., for the $y$ coordinate equal
to $\ldots,n-1,n+1,\ldots$ ($\ldots,n,n+2,\ldots$. The resulting 2D-lattice has
thus the spacing $\lambda/2$ ($\lambda/4$) in the $x$- ($y$-) direction.
The tunnel rates in the $x$ direction are due
to kinetic energy; they are spatially homogeneous and assumed to be equal for
both hyperfine states. The lattice is tilted in the $y$-direction,
which introduces an energy shift $\Delta$ between neighboring columns.
Tilting can be achieved by accelerating the lattice, or by placing it in a
static electric field linearly dependent on $y$.

By this, standard tunneling rates due to kinetic energy are suppressed
in the $y$ direction. Instead, tunneling is laser assisted, and driven
by two pairs of lasers resonant for Raman transitions between
$|g\rangle$ and $|e\rangle$, i.e., $n\leftrightarrow n \pm 1$.
This can be achieved because the offset energy for both
transitions is different and equals $\pm \Delta$.
Detunings of the lasers are chosen in such a way that the effect of tilting is
cancelled in the rotating frame of reference.
The lasers  consist of running waves
in the $\pm x$-direction, so that the corresponding tunneling rates
acquire local phases $\exp(\pm iqx)$.

In order to realize ``artificial'' non-Abelian fields in a similar scheme,
one has to use atoms with internally degenerate Zeeman sublevels
in the hyperfine ground
state manifolds,  $|g_i\rangle$, and $|e_i\rangle$ with $i=1,\ldots, n$,
whose degeneracy is  lifted in external magnetic fields.
These internal states may be thought of as ``colors'' of the gauge fields.
Promising fermionic candidates with these properties are heavy Alkali atoms,
for instance,  $^{40}$K atoms in states $F=9/2, M_F=9/2,7/2, \ldots$,
and $F=7/2, M_F=-7/2, -5/2, \ldots$; in particular, they allow for
realizing ``spin'' dependent lattice potentials
and spin dependent hopping \cite{wilczek}.

Having identified the ``colors'', one should essentially repeat 
the scheme of Ref. \cite{jakschbut} with two
additional modifications: laser assisted tunneling rates
along the $y$-axis should depend on the internal state, although not
necessarily  in the sense of  Ref. \cite{wilczek}. They rather should 
be performed in such a way that for a given link $|g_i\rangle$ to
$|e_i\rangle$, tunneling should be  described by a non-trivial
unitary tunneling matrix $U_y(x)$ being a member of the ``color'' group
($U(n)$, $SU(n)$, $GL(n)$ etc.). For unitary groups,
the tunneling matrix $U_y(x)$ can be represented as
$\exp(i\tilde\alpha A_y(x))$.
Here, $\tilde\alpha$ is real, and $A_y(x)$ is a
Hermitian matrix from the corresponding gauge algebra,
e.g., $\mathfrak{u}(n)$ or $\mathfrak{su}(n)$. Since transitions 
from $|g_i\rangle$ to $|e_i\rangle$ correspond to different frequencies
for each $i$, they are driven by different running wave lasers.
Thus, they may attain different phase factors $\exp(\pm i q_ix)$.
 
\begin{figure}[tbp]
\centering
\caption{{\it Optical lattice setup for U(2) gauge fields:}
Red and blue open semi-circles (closed semi-circles) denote atoms in states
$|g_1\rangle$
and $|g_2\rangle$, respectively ($|e_1\rangle$ and $|e_2\rangle$).
Top) Hopping in the $x$-direction is laser assisted and allows for
unitary exchange of colors; it is described by the same unitary hopping matrix
$U_x$ for both $|g_i\rangle$
and $|e_i\rangle$ states. Hopping along the $y$-direction is also
laser assisted and attains ``spin dependent'' phase factors.
Bottom) Trapping potential in $y$-direction.
Adjacent sites are set off by an energy $\Delta$ due to the lattice
acceleration, or a static inhomogeneous electric field. The
lasers $\Omega_{1i}$ are resonant for transitions $|g_{1i}\rangle
\leftrightarrow|e_{2i}\rangle$, while $\Omega_{2i}$ are resonant for
transitions between $|e_{1i}\rangle
\leftrightarrow|g_{2i}\rangle$ due to the offset of the lattice sites.
Because of the spatial dependence of $\Omega_{1,2}$ (running waves in
$\pm x$ direction)  the atoms hopping around the plaquette get the
unitary transformation $\hat U=U^{\dag}_y(m)U_xU_y(m+1)U^{\dag}_x$,
where $U_y(m)=\exp(2\pi i m \ {\rm diag}[\alpha_1,\alpha_2])$,
as indicated in upper figure.}
\includegraphics[width=0.4\textwidth]{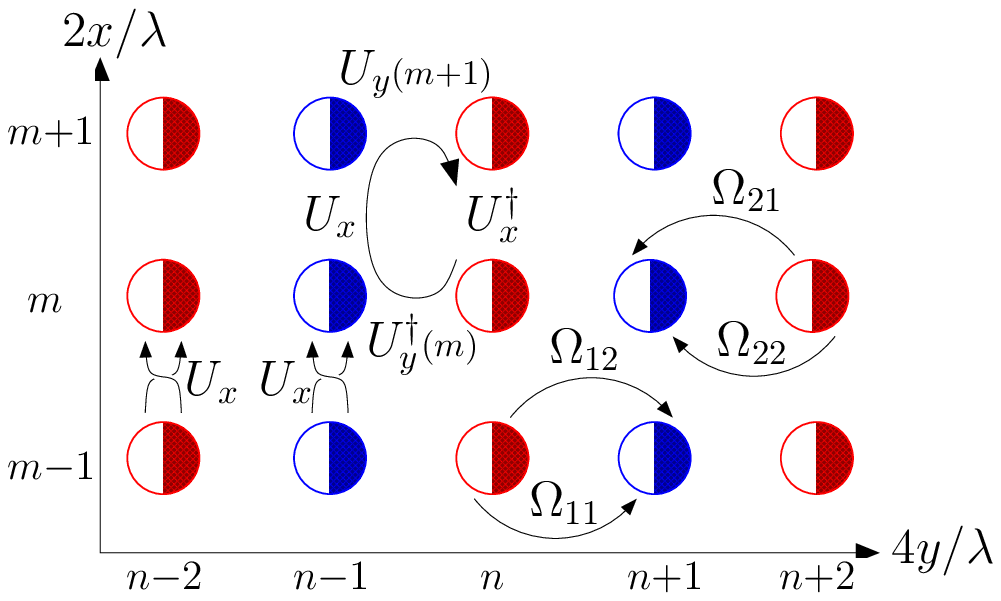}
\includegraphics[width=0.3\textwidth]{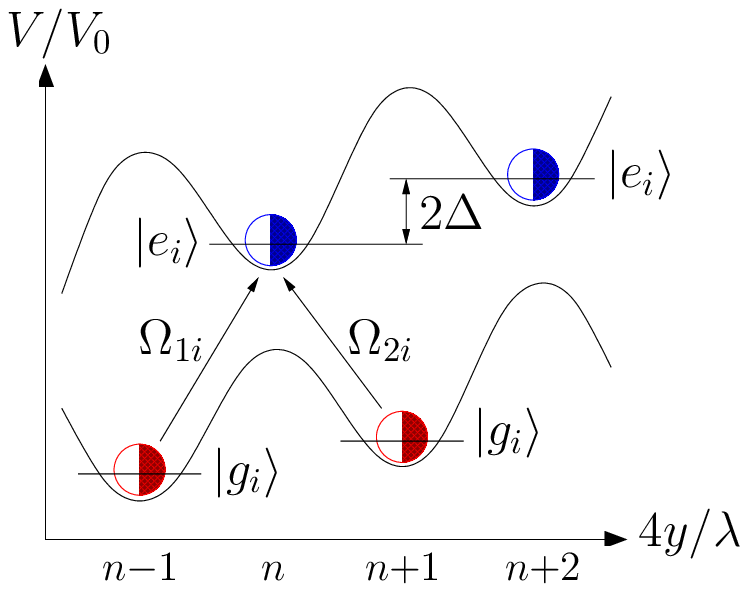}
\label{fig1}
\vspace*{-0.5cm}
\end{figure}

\begin{figure}[tbp]
\centering
\caption{The Hofstadter ``moth'' spectrum.
Eigenenergies $\varepsilon$ are plotted versus
$\alpha_i=p_i/q_i,\in[0,\,=0.5]$ ($i=1,2$), where $q_i\leq 41$
and $\alpha_1\neq\alpha_2$.  }
\includegraphics[width=0.5\textwidth]{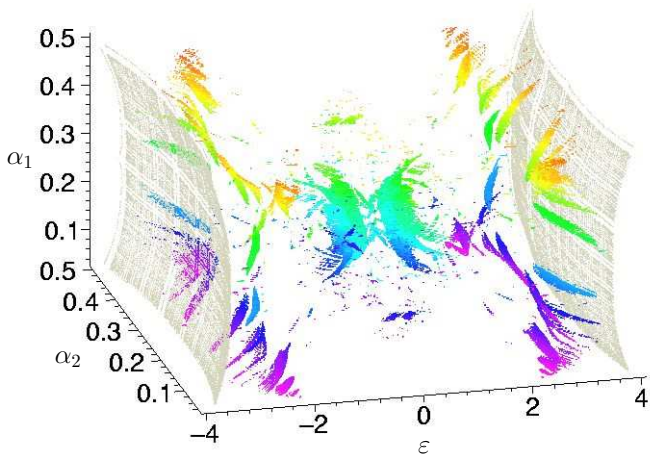}
\label{Hofstaedter}
\vspace*{-1.0cm}
\end{figure}

In order to create gauge potentials that cannot simply be reduced
to two independent Abelian components, tunneling in the
$x$-direction should also be Raman laser assisted and should allow for coherent
transfer between internal Zeeman states. We will assume for now that this is
proceeded by the same unitary tunneling matrix $U_x$ for both hyperfine state
manifolds, although more general situations are feasible
and basically interesting.
To assure a genuine non-Abelian character of the fields,
it is necessary that $[U_x, U_y(x)]\ne 0$.
  We stress that all elements of our scheme,
as  shown in Fig.~\ref{fig1}, are 
experimentally accessible.
Nevertheless, consistent gauge group realizations demand tunneling matrix
amplitudes to be controlled in such
a way that they strictly belong to the corresponding unitary groups.
which assures the vector potentials to be composed of the Hermitian
generators\cite{fotka}.
%
%
The scheme of Fig.~\ref{fig1} allows to generalize the Hamiltonian
(\ref{H}) to the case of non-Abelian
vector potentials. In fact, we replace the components of
$\vec A$  by the corresponding matrices from the
group algebra. In particular, the illustrated setup generates
``artificial'' gauge potentials of the form $\vec A=(A_x,A_y(m),0)$, with
\eqn{\vec{A}(m,n)=\frac{\hbar c}{{\rm e}a}\left(
             \Big(
         \begin{array}{cc}
         0&\frac{\pi}{2}{\rm e}^{\ci\phi}\\
         \frac{\pi}{2}{\rm e}^{-\ci\phi}\hspace*{-0.2cm}&0
         \end{array}\Big),\,
         \Big(
         \begin{array}{cc}
         2\pi m\alpha_1&\!\!\!\!\!0\\
         0&\!\!\!\!\!\!2\pi m\alpha_2
         \end{array}\Big),\,
         0\label{pot}
         \right).}
More precisely, the  scheme associates  a unitary tunneling
operator  with every  link in analogy with
standard lattice gauge theory prescriptions \cite{gauge}:
$U(m-1,n \to m, n)\equiv U_x$, $U(m,n \to m-1, n)\equiv U^{\dag}_x$,
$U(m, n \to m, n+1)\equiv U_y(m)$,
$U(m, n \to m, n-1)\equiv U^{\dag}_y(m)$, where
$U_x=\exp{(+ iea A_x/c\hbar)}$ and $U_y =\exp{(+ iea A_y(m)/c\hbar)}$.
The only difference is that
$\vec{A}$ acquires an overall factor of $\hbar c/ea$;
in effect, though it  does not behave well in the  continuous limit $a\to 0$,
the "magnetic flux" per plaquette, $\alpha_{1,2}$ remains finite.
Thus, we are in the same limit of ultra-intense
fields $\sim 1/a^2$ as in the "classic" Hofstadter case of Abelian
magnetic fields. The ansatz $\psi(ma,\,na)={\rm e}^{\ci\nu n}g(m)$,
leads to a generalized Harper wave equation
\eqn{\left(\!\!
     \begin{array}{c}g({m\!+\!1})\\g({m})\end{array}\!\!\right)
       = B(m)
       \left(\!\!\begin{array}{c}g({m})\\g({m\!-\!1})\end{array}\!\!
       \right)\,\label{GenHarper}}
with       
\eqn{ B(m)\!=\!\left(\!
       \begin{array}{cc}
         \Big(
         \begin{array}{cc}
         0&\hspace*{-0.5cm}{\rm e}^{\ci\phi}\varepsilon_m(\alpha_2,\,\nu)\\
         {\rm e}^{-\ci\phi}\varepsilon_m(\alpha_1,\,\nu)&\hspace*{-0.5cm}0
         \end{array}
             \Big)\hspace*{0.0cm}&\Big(
         \begin{array}{cc}
         \!-1\!&\!0\!\\
         \!0\!&\!\!\!-1\!
         \end{array}\Big)\\\Big(
         \begin{array}{cc}
         \,1&\,0\\
         \,0&\,1
         \end{array}\Big)&\Big(
         \begin{array}{cc}
         \,\,0&\,0\,\\
         \,\,0&\,0\,
         \end{array}\Big)
       \end{array}\!\right)\,,\nonumber}
where $\varepsilon_m(\alpha,\,\nu)=\varepsilon-2\cos(2\pi m\alpha-\nu)$
is the Harper energy term.       
Both, $\psi(m,\,n)$ and $g(m)$ are now two-component objects.
In the particular case of eq.~(\ref{pot}), when
two successive transfer matrices $B(m)$ are multiplied
by each other, eq.~(\ref{GenHarper}) decomposes
into a pair of equivalent 2D equations. Nevertheless, to obtain the spectrum,
one has to rely on numerical methods.
Given each $\alpha_i=p_i/q_i$ rational, the problem is $Q$-periodic
(where $Q$ equals the smallest common multiple of $q_1$ and $q_2$).
The allowed energies are
those for which the product of $Q$ successive matrices $B(m)$
has an eigenvalue of modulus 1. The spectrum shows a band structure,
is continuous, and is bounded by two hyperplanes (depicted in grey in
Fig.~\ref{Hofstaedter}).
It exhibits a very complex formation of holes of finite
measure and various sizes, which we name the Hofstadter ``moth''. 
Although a rigorous proof can not be provided, the ``moth'' reminds of a
fractal structure. Obviously, this fractal structure will
be very sensitive to any sort of perturbation
(finite size of the system, external trapping potential etc.)
on very small scales.
But, since the holes are true 3D objects with finite volume, 
the spectrum will be more robust on a larger scale
to perturbations than in case of the Hofstadter ``butterfly''.
This may be revealed in comparison of our ``moth'' with two uncoupled
Abelian ``butterflies''. The latter would form a spectrum of allowed energies
consisting of intersecting lamellas of zero width. Here, the perturbations
are required to guarantee allowed regions of finite volume.
Since $\alpha_1$ and $\alpha_2$ are easily controllable, the spectral
structure is directly measurable; a possible scheme would be to load
a dilute (weakly interacting, or practically non-interacting) Bose condensate
into the lattice and measure its excitations, e.g., by looking at the
time evolution of the particle density, as suggested in Ref.~\cite{jakschbut}. 
Alternatively, one could load an ultra-cold polarized Fermi gas and measure
its Fermi energy as a function of the number of particles.

It is interesting to consider yet another effect that becomes particularly
spectacular in the limit of ultra-intense gauge fields,
and that can be measured in the proposed system: a non-Abelian
Aharonov-Bohm effect, to be considered as an example of non-Abelian
interference. In order to realize it,
one should prepare, for instance, a weakly interacting Bose condensate
in a definite internal state $|\psi_0\rangle$ around a location $P_1$.
Then, the BEC (or parts of it) should be split by Raman scattering,
and piecewise dragged (using, e.g, laser tweezers) to a meeting point $P_2$
on two distinct paths. These correspond to the unitary transporters
$U_1$ and $U_2^\dag$, respectively.
A measurement of the density of atoms at $P_2$ will reveal a non-Abelian
interferometer signal,  i.e., it will detect the interference
term $n_{int}\propto \langle \Psi_0|U_2U_1|\psi_0\rangle$.
Choosing, e.g., the rectangular loop, consisting of $L_{x_1}$ steps
in the $-x$-direction, $L$ steps in the $y$-direction, $L_{x_1}+L_{x_2}$
steps in $x$-direction, $L$ steps in $-y$ directions, and finally $L_{x_2}$
steps in $-x$-direction,  we obtain for the gauge potential of eq.~(\ref{pot}): 
$ n_{int} \propto 1 $ if $L_{x_1}+L_{x_2}$ is even, and
$n_{int}\propto \langle\Psi_0|
 \exp{[\pm 2\pi i\hat\alpha L(L_{x_1}+L_{x_2})]}
 |\psi_0\rangle$, if $L_{x_1}+L_{x_2}$ is odd,
where $\hat \alpha={\rm diag}(\alpha_1, \alpha_2)$.
The signal is thus extremely sensitive to $L_{x_1}$ and $L_{x_2}$.
If one attempted to measure the phase shifts by introducing  obstacles
on the $y$ arms of this interferometer, the result would
strongly depend on the $x$-coordinate of the obstacles,
and the location of $P_1$, $P_2$; a non-Abelian manifestation of the external
gauge potential.

Obviously, the properties of the considered system in the limit of
ultra-intense fields are quite complex. Though, to get a better intuition
concerning the scope of this scheme, it is useful to consider
also the ``continuum'' limit $a\to 0$ with $V_0\to a^2/m$.
Then, the Hamiltonian becomes ${\cal{H}}=(\vec p-e/c\vec A)^2/2m$.
A natural question to ask is what kinds of gauge fields of ``normal'',
i.e., $a$-independent strength, can be realized?
In other words: what are the possible ``artificial'' gauge potentials that
can be created? In general, phase factors resulting from running
wave vectors can be introduced for all tunneling matrices:
$\vec{A}(\vec r)= \frac{c}{{\rm e}l}(M_1 + [M_2
(\frac{x}{l})+M_3(\frac{y}{l})], N_1 + [N_2
(\frac{x}{l})+N_3(\frac{x}{l})], 0)$, where $M_i, N_i$ are arbitrary
(in general non-commuting), dimensionless, and $a$-independent matrices
from, e.g., $\mathfrak{u}(2)$, and $l$ is the characteristic length on
which  $\vec A$ varies.
Note that in order to realize non-Abelian interferometry,
it is sufficient to fulfill $[M_1,N_1]\ne 0$ and set all other
matrices to zero.
This may be achieved using simpler experimental arrangements
as those discussed above, and will be the subject of further
investigation.
Furthermore, local disorder may be introduced in a controlled way
that allows for small fluctuations of the matrices $M_i$. In particular,
disorder can be made annealed, i.e. it changes on a time scale comparable
with the relevant time scales of the system, and, thus, mimics thermal
fluctuations. It can be of significant amplitude,  provided it does not
drive the assisting lasers system out of resonance.
For example, wave vectors of running waves can strongly fluctuate and force
the magnetic fluxes to fluctuate correspondingly.
Even more complicated spatial dependences, e.g., piecewise linearity,
of $\vec A$ are feasible by using static electric fields, laser induced
potentials, etc. Additional lasers may introduce local,
and in general time dependent  unitaries.
Such transformations would generate arbitrary local temporal
components of the gauge potential, $A_0(x,\,y)$.
Although for Yang-Mills fields in 2+1D, this component may be gauged out
adapting the Weyl, or strict temporal gauge \cite{schulz}, the corresponding
gauge transformations may introduce more complex spatial and temporal forms
of the remaining two components of $\vec A$. 

Finally, let us discuss, in which sense the proposed 2D scheme might be useful
to study lattice gauge theories (LGT) in 2+1D. There, one uses the framework
of Euclidean field theory, and various methods
of statistical physics, such as Monte Carlo methods, to sum over all
configurations of gauge and matter fields at finite or infinite
temperatures. Thus, gauge fields are dynamical variables in LGT,
but are obviously not in the proposed scheme.
Moreover, our scheme is realized in real rather than in imaginary time,
thus, only a limited set of configurations of gauge fields can be generated.
Nevertheless, the big advantage of our proposal is that given a gauge
field configuration, the dynamics of matter fields in real time are
given for free. When experimentally realized, the system may be used as a
quantum simulator of matter fields in a given external gauge field.
By generating  other configurations, we may try to
"mimic" the Monte Carlo sampling of LGT. Averaging over both, annealed
disorder and quantum fluctuations should approximate the statistical
average in LGT. However, this inevitably requires that
generated configurations represent the characteristic
or statistically relevant ones of corresponding LGT phases.

Apparently, with the limited gauge field configurations accessible in the
proposed scheme, not too much can be done. Anyhow, at least some configurations
may be generated that share characteristics of LGT phases, e.g.,
an area law fulfilled by Wilson loops in the confinement sector.
In fact, the $SU(2)$ gauge potential of eq.~(\ref{pot}) with fluctuating,
but anticorrelated fluxes $\alpha_1=-\alpha_2$ yields an area law for Wilson
loops in the $xy$-plane, provided the probability distribution of the fluxes
is Lorentzian.
However, it would be desirable to create configurations that exhibit other
characteristics of the confinement phase such as appropriate distributions
of center vortices, Abelian magnetic monopoles, instantons, merons, calorons
etc.~\cite{equiv,engel}. 

We expect  that this program will lead to many fascinating results.
It should be stressed that Yang-Mills theories in 2+1D are in the center
of interest in high energy physics, as they describe the high temperature
behavior of 4D models \cite{jackiw}. Recently, there has been
progress in understanding these theories \cite{schulz,kkn} in the
continuum limit, and in the pure gauge sector: the gauge-invariant degrees
of freedom have been identified and the Hamiltonian has been constructed.
The ground state wave function is, to a very good approximation, a Gaussian
function of the currents,  and quantities as the string tension
are known exactly. Our lattice models may shed new light on these
recent discoveries.

\begin{acknowledgments}
We acknowledge fruitful and inspiring discussions with
H.-J. Diehl, O. Lechtenfeld, H. Frahm, F. Haake, C. S\"{a}mann,
and H. Schulz. This work was supported by the Deutsche Forschungsgemeinschaft
(SPP1116, SFB407, 436POL), the ESF PESC QUDEDIS, and
the Spanish MEC grant HA2002-0036.
\end{acknowledgments}


\begin{thebibliography}{999}
\bibitem{fotka} In general tunneling matrices could belong to $GL(n)$;
such models are of interest themeselves.
\end{thebibliography}
\end{document}